# Electrical and photoelectrical characteristic investigation of a new generation photodiode based on bromothymol blue dye

A Gencer Imer[1], A Tombak[2] and A Korkut[1]

[1]Department of Physics, Yuzuncu Yil University, 65080, Van, Turkey
[2]Department of Physics, Batman, University, Batman, Turkey

E mail: gencerimerarife@gmail.com

**Abstract.** Bromothymol blue (BTB) with the molecular formula of $C_{27}H_{28}Br_2O_5S$ was grown onto p-Si substrate to fabricate heterojunction by spin coating technique. The current voltage (*I-V*) measurements of diode were carried out in dark and under different illumination intensity at room temperature. The photoelectrical properties of heterojunction based on BTB were investigated using the illumination intensity dependent *I-V* data. The results showed that photo current of diode increases with the increase in light intensity. Also, the electrical parameters of device were determined via *I-V,* and capacitance- voltage (*C-V*), conductance-voltage (*G-V*) measurements at different frequencies. It is observed that the excess capacitance is created at low frequencies due to the contribution of interface states charge which can follow the alternative current signal to capacitance. It is stated that, both the electrical & photoelectrical parameters of diode can be changed, and also the performance of the device could be affected by the organic thin film interlayer.

1. **Introduction**

The organics or molecular materials have a wide variety of applications in modern electronic technology owing to structural flexibility, lower production cost, easy of synthesis and the possibility of using for large area device production [1,5]. Also, the organic thin films can be easily deposited onto different kind of substrates via deep coating [6], spray pyrolysis [7], spin coating [8-13], which are very common and cost effective growing methods. Namely, electronic and optoelectronic properties of organic/inorganic semiconductor devices have been studied as Schottky didoes [5-7], photosensor [13], photodiodes [6,12], and solar cells [10-15] by a great number of authors. Kılıcoglu et.al [10] fabricated the *TFF/p-InP* heterojunction, and investigated the electrical and photoelectrical properties of that device. It was found that this device can be used in the photovoltaic applications, because of the strong light response. Ocak [16], and Güllü [17,18] have reported the value of barrier height for the organic inorganic heterojunction devices can be modified by the organic interlayer at the MS interface owing to the tunneling effect through the organic interlayer.

In the present study, we have tried to study rectifier and photoresponce behavior of the bromothymol blue/p-Si (organic/inorganic semiconductor) contacts. The aim of this study to fabricate an *Al/BTB/p-Si* structure and to investigate the electrical and photoelectrical properties of device via current voltage (*I-V*) measurement in dark and under different illumination intensities; and capacitance-voltage (*C-V*), conductance-voltage (*G-V*) measurements at different frequencies. Both the fundamental photoelectrical parameters and electrical parameters such as ideality factor (*n*), the barrier height ($\Phi_b$) and series resistance ($R_s$) are evaluated by different methods.

## 2. Experimental procedure

For the fabrication of *Al/BTB/p-Si* heterojunction, p-Si substrate (with (100) orientation and 1-10 Ω-cm resistivity) was chemically cleaned via RCA method. For the formation of ohmic contact, the aluminum metal was evaporated to substrate; the obtained structure was annealed at 570 °C for 3 min in $N_2$ atmosphere. For the synthesis of BTB thin films, bromothymol blue was dissolved in methanol, and the precursor solution of $1\times10^{-2}$ molL$^{-1}$ BTB was prepared at fist. The precursor solution was coated onto the front side of silicon substrate with a rotating speed of 2000 rpm for 1 min by spin coater. Later, top contacts were formed onto the obtained BTB/p-Si structure using through a shadow mask with the shape of circular dots of 1.5 mm in diameter. The contacts having were labeled as D1, D2, D3. The *I-V* measurements of diode were done at room temperature using Keithley 2400 sourcemeter in dark and under a Newport 96000 solar simulator with AM 1.5 global filter with different illumination intensities. The *C-V, G-V* measurements at different frequencies were carried out by means of HP Agilent 4294A impedance analyzer.

## 3. Results and discussions

Current-voltage (I-V) measurements of the *BTB*/p-Si contacts were carried out in dark at room temperature to investigate the electrical properties of them. Semi-logarithmic *I-V* plots of *Al/BTB/p-Si* heterostructure for each diode have been represented in Figure 1. The structure has a good rectifying behaviour.

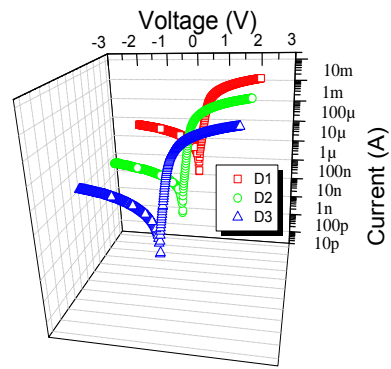

**Figure 2.** Experimental forward and reverse bias current versus voltage characteristics of the *Al/BTB/p-Si* structure for different dots

Thus, it can be said that organic-inorganic heterojunction reveals the property of Schottky like MS junction, and thermionic emission (TE) theory can be used to extract electrical parameters. According to the theory, the net current with series resistance ($R_s$) can be expressed as [19],

$$I = I_0 \exp\left[\left(\frac{q(V-IR_s)}{nkT}\right) - 1\right] \quad (1)$$

where *q* is the electronic charge, *V* is the applied voltage, *k* is the Boltzmann constant, *T* is the ambient temperature in Kelvin, and $I_0$ is the saturation current obtained by extrapolation of the linear portion of semi-log *I-V* plot to zero applied voltage and is defined as;

$$I_0 = AA^*T^2 \exp\left(-\frac{q\phi_b}{kT}\right) \tag{2}$$

here, $\phi_b$ is the zero bias barrier height of device, $A$ is the effective contact area (0.018 cm$^{-2}$), $A^*$ is the Richardson constant and is equals to 32 Acm$^{-2}$K$^{-2}$ for p-Si [20]. $n$ is an ideality factor which is a dimensionless quantity. It can be obtained from the slope of the linear portion of the forward bias semi-log *I-V* plot through the relation:

$$n = \frac{q}{kT}\frac{dV}{d\ln(I)} \tag{3}$$

Furthermore, the value of the barrier height of the device can be obtained as,

$$\phi_b = \frac{kT}{q}\ln(\frac{AA^*T^2}{I_0}) \tag{4}$$

The values of the barrier height and ideality factor of the fabricated structure were determined and given in Table 1.

The mean value of barrier height and ideality factor was evaluated as 0.729 eV and 1.41 using Eqs. (3) and (4), respectively. $\phi_b$ value of heterojunction is 0.729 eV, which is 0.149 eV higher than that of conventional Al/p-Si MS contact (0.58 eV) [21]. This enhancement may be derived from the saturation of dangling bonds at the surface of the inorganic semiconductor by the bromothymol blue thin film [22]. The modification in that of the barrier height by the organic interfacial layer has been studied in the many of research effort [7-13, 21-28]. For instance, the quantity of increment of 0.19 eV in the value of energy barrier was reported by Ugur et al [12] for the Al/p-Si heterojunction with quinoline yellow interfacial layer.

In addition, the mean ideality factor of heterostructure was found to be as 1.41, (n>1). So, the diode deviates from ideality because of organic interface layer at the interface. The higher ideality factor value of diode may be originated from the interface state density at the interface, existence of series resistance and presence of native oxide at interface [21-27]. Deviation from the linearity at high forward bias voltages in semi-log *I-V* plot is giving rise to the effect of series resistance, $R_s$. In order to determine of the $R_s$, modified Norde' function [28] can be used as follows:

$$F(V) = \frac{V}{\gamma} - \frac{kT}{q}\ln\left(\frac{I(V)}{AA^*T^2}\right) \tag{5}$$

where $\gamma$ is the first integer greater than *n*, ( *γ = 2 herein*). F(V) vs. V curve of the Al/BTB/p-Si device is shown in Fig.3. The barrier height values of heterojunction can be obtained by using the equation,

$$\phi_b = F(V_{min}) + \frac{V_{min}}{\gamma} - \frac{kT}{q} \tag{6}$$

where $F(V_{min})$ is the minimum value of $F(V)$ and $V_{min}$ is the corresponding voltage value. The values of $F(V_{min})$ and $V_{min}$ are determined, and listed in the Table 1 for different contacts. In this method, the series resistance $R_s$ can be calculated through the relation,

$$R_s = \frac{kT(\gamma - n)}{qI_{min}} \qquad (7)$$

where $I_{min}$ is the corresponding current value at the $V_{min}$. The values of the barrier height and series resistance of each diodes for *Al/BTB/p-Si* heterojunction were calculated, and also listed in Table 1, the average value of series resistance for that heterojunction was extracted as as 0.771 eV and 980 Ω by using Eqs (6) and (7), respectively. The value of barrier height obtained from Norde' function is very close to that of obtained forward bias semi-log *I-V*.

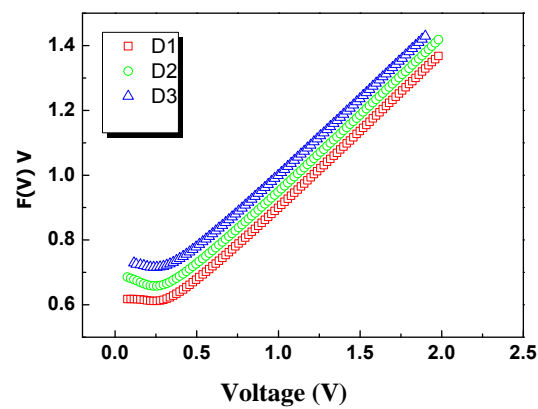

**Figure 3.** Experimental *F(V)* vs. *V* plots for different dots of *Al/BTB/p-Si* structure at the dark

**Table 1.** Electrical characteristic parameters for different contacts of *Al/BTB/p-Si* structure

| ID | From I-V plot | | From Norde plot | | | |
|---|---|---|---|---|---|---|
|  | n | $\Phi_b$ (eV) | $F(V)_m$ (V) | $V_{min}$ (V) | $R_s$ (kΩ) | $\Phi_b$ (eV) |
| D1 | 1.3 | 0.75 | 0.662 | 0.276 | 0.74 | 0.77 |
| D2 | 1.4 | 0.72 | 0.658 | 0.236 | 1.25 | 0.75 |
| D3 | 1. | 0.72 | 0.671 | 0.306 | 0.95 | 0.79 |

The effect of illumination on the *I-V* characteristic of the heterojunction was carried out and seen in Figure 4. The reverse bias current increases extremely with the increasing light intensity, while forward bias current of that is almost constant. This indicates the generated electron-hole pairs in organic-semiconductor junction by incident photons. Photocurrent is produced by photo generation of the free carriers contributing to the reverse current after the process of absorption of photon energy at the junction. Furthermore, photocurrent and the photovoltage are increased with the increasing light intensity as seen from this figure. The characteristic photovoltaic parameters such as short circuit current $I_{SC}$, open circuit voltage $V_{OC}$ were defined as 54.68 µA and 406 mV under 100 mW/cm$^2$ illumination, respectively. These characteristic parameters of *Al/BTB/p-Si* structure under different illumination intensities were listed in the Table 2.

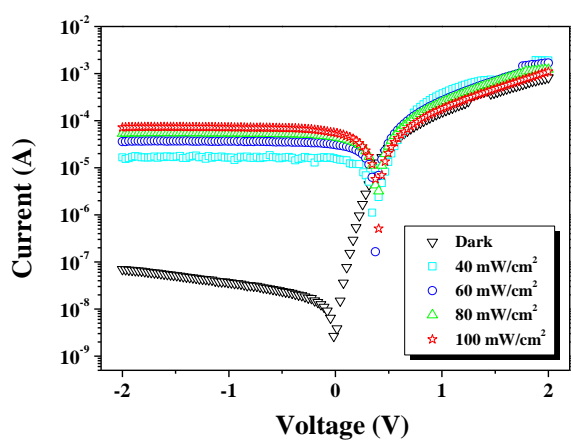

**Figure 4.** Semilogaritmic *I-V* plots of *Al/BTB/p-Si* at the dark and under different illumination intensities

The power conversion efficiency (PCE) was evaluated, and given in the Table 2. The obtained characteristic parameters of studied device are poor for the solar cell application, but suitable for photodiode, photosensor application [30,6].

In order to determine the phoconductive responsivity of the diode to the different illumination intensity, the following equation can be used [20,31];

$$R = I_{PH}/L.A \qquad (8)$$

where, *A* is the area, which is exposed to illumination. The calculated values of *R* were given in Table 2. It is understood from table that these values exponentially increase with incident light intensity. This means that the responsivity to the light for this diode is dependent to the illumination power which is resulted in the generated electron-hole pairs by photons.

**Table 2.** The values of $V_{OC}$, $I_{SC}$, *PCE*, *R* at different light intensity

| Light Intensity (mW/cm$^2$) | $V_{OC}$ (mV) | $I_{SC}$ (µA) | PCE | Responsivity, R (A/W) (x10$^{-4}$) |
|---|---|---|---|---|
| 40 | 347 | 14.69 | 0.38 | 3.96 |
| 60 | 375 | 30.69 | 0.46 | 5.69 |
| 80 | 395 | 44.28 | 0.49 | 6.36 |
| 100 | 407 | 54.78 | 0.43 | 6.79 |

*C-V* measurements of *Al/BTB/p-Si* heterojunction were carried out in the range from 100 to 500 kHz frequencies. The Figure 5 presents the relation of capacitance with the applied voltage at different frequency for the studied photodiode. It is seen from the figure that the capacitance of diode is almost constant up to the voltage of ~ 1 V, reaches to the maximum value, and forms a peak owing to the interface states at the interface of the device. The maximum value of *C-V* plots increases at low frequency regime, as a result of that the interface states charges could follow the alternating signal (a.c.) [10,15].

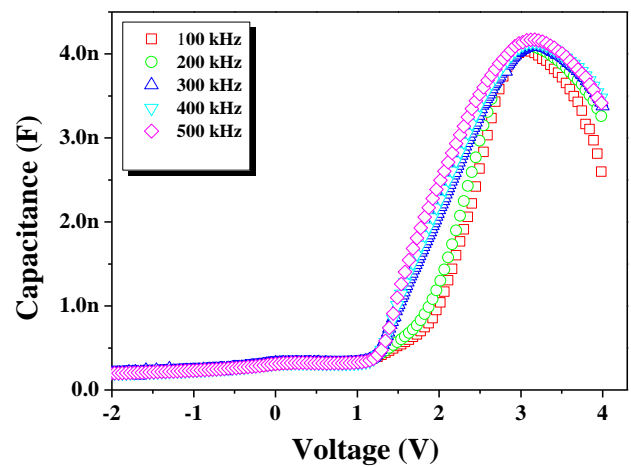

**Figure 5.** Capacitance vs. voltage plots of *Al/BTB/p-Si* structure under different frequencies.

In addition, the capacitive property of the diode can be analyzed as a function of reverse bias voltage. Thus, the depletion region capacitance of device can be described as [20],

$$\frac{1}{C^2} = \frac{2(V_{bi} + V)}{q \varepsilon_s \varepsilon_0 A^2 N_a} \qquad (9)$$

where $V_{bi}$ the built in potential, which is estimated from the intersection of linear portion of the reverse bias $C^{-2}$-$V$ plot, $\varepsilon_S$ is the dielectric constant of semiconductor (=11.7 for Si), $\varepsilon_0$ = 8.85 x $10^{-14}$ F/m, $A$ is the effective diode area. $N_a$ is the density of ionized acceptors, which is found from the slope of linear portion of the same plot. The Figure 6 shows the $C^{-2}$-$V$ plot of *Al/BTB/p-Si* structure for 500 kHz, as a representative example. The values of $V_{bi}$ and $N_a$ were calculated as 0.829 V and 4.81x$10^{15}$ cm$^{-3}$ to the $V$ axis, respectively. The barrier height value can be obtained from the relation;

$$\phi_{b(C-V)} = eV_{bi} + kT \ln(\frac{N_v}{N_a}) \qquad (10)$$

The second term in the Eq. (10) is known as the potential difference between the top of the valance band in the neutral region of p-Si, in which $N_v$ = 1.04×$10^{19}$ cm$^{-3}$. Therefore, the $\phi_b$ value was calculated as 1.028 eV using Eq. (10) from reverse bias $C^{-2}$-$V$ characteristic of the heterostructure.

This barrier height value is higher than that of obtained from the *I-V* data. The reason of this difference can be the presence of trap states and interfacial layer, interfacial layer, and oxide layer at interface of the device [31,32].

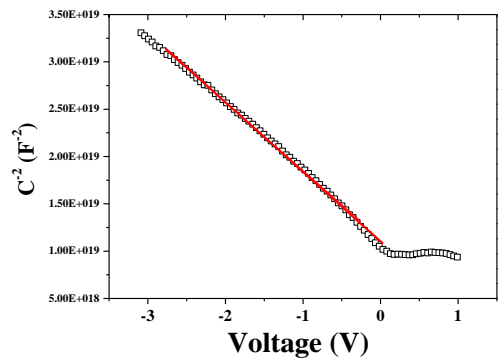

**Figure 6.** $C^{-2}$ -V plot of *Al/BTB/p-Si* structure for 500 kHz

The conductance (*G)* was measured, and the Figure 7 exhibits (*G-V*) characteristics of *Al/BTB/p-Si* heterojunction at different frequencies. It is seen from the figure the value of conductance for the studied diode increases as the frequency increases for the forward bias voltage, while that remains almost constant for the reverse bias voltage. Such behaviour in the *G-V* characteristics of specimen may be ascribed to the presence of series resistance, and very thin insulator layer on the semiconductor [6,10,11,30].

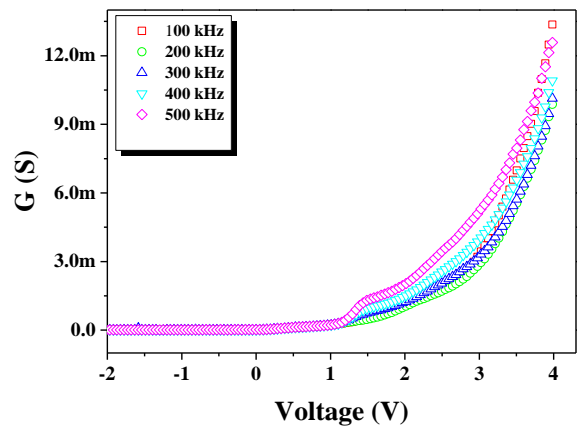

**Figure 7.** Conductance vs. voltage plots of *Al/BTB/p-Si*
 structure at different frequency

In order to evaluate the density of interface states ($D_{it}$) profile as a function of frequency, the reliable method is developed by Hill-Coleman [33], according to this method, $D_{it}$ can be estimated by the following relation,

$$D_{it} = \frac{2}{qA} \frac{G_m/\omega}{[\left(\frac{G_m}{\omega C_{ox}}\right)^2 + (1 - \frac{C_m}{C_{ox}})^2]} \tag{11}$$

where $G_m$ is the maximum conductance obtained from the $G$-$V$ plot, $\omega$ is the angular frequency, $C_m$ is the capacitance value corresponded to the $G_m$, $C_{ox}$ is the capacitance of the interface layer in accumulation regime. The value of $D_{it}$ diminishes with the increasing frequency, and varies from ~3.5 x $10^{11}$ to 5.9 x $10^{10}$ eV$^{-1}$cm$^{-2}$ as seen in the Figure 8 for the studied heterojunction. This value for the studied hetero structure is lower than both the that of conventional Al/p-Si diode (~$10^{13}$ eV$^{-1}$cm$^{-2}$ at room temperature) [34], and that of MIS structures reported in the literature [7,11,24]. Thus, both the electrical & photoelectrical parameters of diode can be modified, and also the performance of the device can be improved by the organic thin film interlayer.

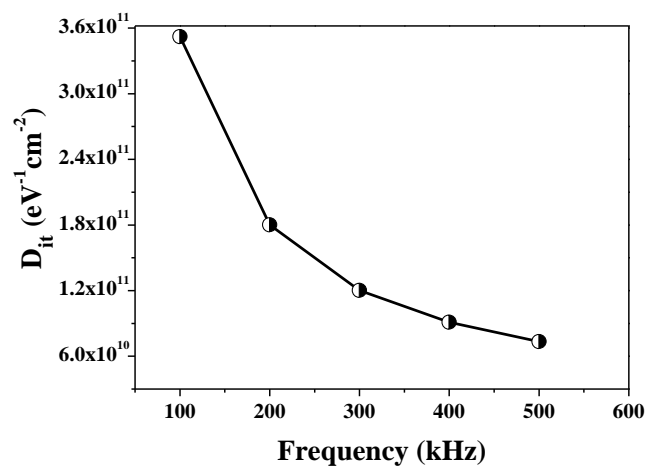

**Figure 8.** Density of interface states vs. frequency plot of *Al/BTB/p-Si* structure

## 4.Conclusion

In this work, *Al/BTB/p-Si* heterojunction was fabricated by forming the *BTB* organic thin film on the p- type silicon crystal using spin coating technique. The photoelectrical characteristics of heterojunction were investigated under different illumination intensities. The photocurrent and photovoltage of the diode increase with the illumination intensity. Also, photoconductive responsivity increases with incident light power. The value of *s* is found as 0.614 for the diode, which suggest continuous distribution of localised interface states. The characteristic photovoltaic parameters such as $I_{sc}$, $V_{oc}$ were determined for different illumination. The obtained results conclude that the forming of the *BTB* organic thin interlayer promotes the photoconductivity due to generation of the electron-hole pairs. It is good candidate for the photo sensor and photo diode applications. Furthermore, the main electrical characteristic parameters for different contacts of *Al/BTB/p-Si* diode, were determined using *I-V,* and frequency dependent *C-V,* and *G-V* measurements at room temperature. $D_{it}$ value of diode is significantly lower in comparison with that of commercial Al/p-Si diode. The obtained results confirm that the electrical properties of the heterojunction can be improved using the interface layer based on organic dye.

**Acknowledgements:** This work has been partially supported by the Yuzuncu Yil Scientific Research management (YYU-BAPB) projects under the contract 2012-FED-B007, 2015-FBE-YL007 and 2015-FBE-YL347. We would like to also thank the Science and Technology Application and Research Center in Dicle University (DUBTAM) for their support.